\documentclass{jfm}

\usepackage{graphicx}
\usepackage{natbib}

\ifCUPmtlplainloaded \else
  \checkfont{eurm10}
  \iffontfound
    \IfFileExists{upmath.sty}
      {\typeout{^^JFound AMS Euler Roman fonts on the system,
                   using the 'upmath' package.^^J}%
       \usepackage{upmath}}
      {\typeout{^^JFound AMS Euler Roman fonts on the system, but you
                   dont seem to have the}%
       \typeout{'upmath' package installed. JFM.cls can take advantage
                 of these fonts,^^Jif you use 'upmath' package.^^J}%
      }
  \else
  \fi
\fi

\ifCUPmtlplainloaded \else
  \checkfont{msam10}
  \iffontfound
    \IfFileExists{amssymb.sty}
      {\typeout{^^JFound AMS Symbol fonts on the system, using the
                'amssymb' package.^^J}%
       \usepackage{amssymb}%
         \let\leq=\leqslant
       \let\ge=\geqslant  \let\geq=\geqslant
      }{}
  \fi
\fi

\ifCUPmtlplainloaded \else
  \IfFileExists{amsbsy.sty}
    {\typeout{^^JFound the 'amsbsy' package on the system, using it.^^J}%
     \usepackage{amsbsy}}
    {\providecommand\boldsymbol[1]{\mbox{\boldmath $##1$}}}
\fi

\providecommand\bnabla{\boldsymbol{\nabla}}

\newcommand\Rey{\mbox{\textit{Re}}}  

\newsavebox{\astrutbox}
\sbox{\astrutbox}{\rule[-5pt]{0pt}{20pt}}

\title[A reduced model of pipe flow]
{Turbulent dynamics of pipe flow captured in a reduced model: puff
relaminarisation and localised `edge' states}

\author[A. P. Willis and R. R. Kerswell]%
{
A\ls S\ls H\ls L\ls E\ls Y\ns P.\ns W\ls I\ls L\ls L\ls I\ls S
\thanks{Present address : Laboratoire d'Hydrodynamique, Ecole Polytechnique, 
91128 Palaiseau, France}\break
\and 
R\ls I\ls C\ls H\ns  R.\ns  K\ls E\ls R\ls S\ls W\ls E\ls L\ls L
}

\affiliation{School of Mathematics,
   University of Bristol,
   BS8 1TW Bristol, United Kingdom
}

\pubyear{2007}
\volume{1}
\pagerange{1--2}
\date{?? and in revised form ??}

\usepackage{epsfig,psfrag}
\usepackage{latexsym}

\renewcommand{\vec}[1]{\mbox{\boldmath $#1$}}
\newcommand{\vechat}[1]{{\skew3\hat{\vec{#1}}}}

\newcommand{\laplace}{\nabla^2}  

\newcommand{\grad}{\bnabla}

\newcommand{\curl}{\bnabla \wedge}

\newcommand{\pd}[1]{\partial_{#1}}
\renewcommand{\vec}[1]{\mbox{\boldmath $#1$}}

\begin{document}

\maketitle

\begin{abstract}
Fully 3-dimensional computations of flow through a long pipe demand
a huge number of degrees of freedom, making it very expensive to
explore parameter space and difficult to isolate  the structure of
the underlying dynamics. We therefore introduce a `2$+\epsilon$'
dimensional model of pipe flow which is a minimal
3-dimensionalisation of the axisymmetric case: only sinusoidal
variation in azimuth plus azimuthal shifts are retained, yet the
same dynamics familiar from experiments are found. In particular the
model retains the subcritical dynamics of fully resolved pipe flow,
capturing realistic localised `puff'-like structures which can decay
abruptly after long times, as well as global `slug' turbulence.
Relaminarisation statistics of puffs reproduce the memoryless
feature of pipe flow and indicate the existence of a Reynolds number
about which lifetimes diverge rapidly, provided that the pipe is
sufficiently long. Exponential divergence of the lifetime is
prevalent in shorter periodic domains.  In a short pipe, exact
travelling-wave solutions are found nearby to flow trajectories on
the boundary between laminar and turbulent flow. In a long pipe, the
attracting state on the laminar-turbulent boundary is a localised
structure which resembles a smoothened puff. This `edge' state
remains localised even for Reynolds numbers where the turbulent
state is global.
\end{abstract}


\section{Introduction}

Laminar flow through a pipe is possible under controlled laboratory
conditions up to flow rates well beyond those at which turbulence is
typically observed. \cite{pfenniger61} achieved laminar flow at
Reynolds numbers, $\Rey := UD/\nu$, as high
as $100\,000$, where $U$ is the mean axial speed, $D$ the diameter
and $\nu$ the kinematic viscosity, indicating that,
rather than the transition to
turbulence being via a linear instability, some other mechanism must
be responsible. Given an initial disturbance of sufficiently large
amplitude, self-sustained turbulence is observed for $\Rey$ of
approximately 2000. This turbulent flow exhibits distinct spatial
structures at different flow rates. For $\Rey$ up to around 2250 the
region of turbulence remains localised, with a length of
approximately $20\,D$, and is referred to as a `puff'
\citep{wygnanski73}. At larger flow rates these puffs slowly
delocalise by splitting into two or more puffs. At much larger
$\Rey$, of around 2800, the disturbances develop into a rapidly
expanding active region of turbulence, referred to as a `slug'. No
explanation has been offered that predicts such a progression in
flow regimes and many issues remain unresolved.

The dynamics of perturbations at transitional Reynolds numbers is
believed to be strongly influenced by a rapidly increasing number of
branches of exact solutions that have been found to appear at these
$\Rey$ \citep{faisst03,wedin04,kerswell05,pringle07,pringle08}. At
$\Rey < 1750$, puffs are observed to suddenly and unexpectedly decay
in experiments \citep{peixinho06}, and it has been suggested that
the turbulent state wanders between these unstable solutions before
relaminarising \citep{hof04,faisst04}. 
The same data also suggest that the mean lifetime for a
puff becomes infinite at $\Rey=1750$ indicating that the puffs
become permanent states at this point \citep{peixinho06}. This
critical $\Rey$ has been reproduced to within 7\% using numerical
computations which adopted the experimental protocol for initiating
the puffs and worked within a long periodic pipe of $50\, D$, so as
to realistically capture the puff structure \citep{willis07}.
However, experiments using a different way of initiating the puffs
and designed to capture longer puff transients claim that no such
critical $\Rey$ exists \citep{hof06}. Simulations also presented
there in a short $\approx 5 \, D$ pipe appear to support this
conclusion. The obvious question is then whether
numerically-simulated turbulence which fills a short pipe has the
same relaminarisation characteristics as localised puff turbulence
captured in longer numerical domains. A complete statistical study
using fully-resolved 3-dimensional computations across a spectrum of
periodic pipe domains remains prohibitively expensive, whereas a
survey using a realistic model system could provide a clarifying
demonstration of difference.

Evidence has also emerged recently that many of
the exact solutions known thus far sit on a separatrix between
laminar and turbulent states, forming an `edge' to the chaotic
region of phase space
\citep{schneider07b,kerswell07,duguet07,willis08} 
(see \cite{itano01,wang07,viswanath07} for similar
observations in channel and plane Couette flow). \cite{schneider07b}
have examined the dynamics of flow restricted to lie in this
separatrix in a short $5 \,D$ pipe, finding at long times a chaotic
attractor apparently centred on a simple travelling wave solution
\citep{pringle07}. In such a short pipe, the turbulence naturally
fills the pipe when triggered and the laminar-turbulent boundary
end-state, or `edge' state, is also a global state. In a longer
($\geq 25D$) pipe, however, localised puffs are the naturally
triggered state at low $\Rey$ which raises the issue of what the
corresponding `edge' state is and how it varies with $\Rey$. For
example, is it initially localised and does it lose localisation at
the same $\Rey$ as the turbulent puff? Again, a realistic model
system can suggest probable answers to these questions quickly.

The use of model systems is well established in plane Couette flow,
which exhibits the same abrupt subcritical transition behaviour as
pipe flow. Several approaches have been designed to reduce the
number of degrees of freedom of this problem in order develop more
tractable models.  The minimal flow unit introduced by
\cite{jimenez91} has been useful in identifying the key components
that lead to self-sustaining turbulence in a very small domain
\citep{hamilton95}.  The model by \cite{lagha07} severely truncates
the degrees of freedom in the cross-stream direction, but captures
spanwise and streamwise spatial structures observed in plane Couette
flow. Using this model, attempts have been made to measure the
lifetime of localised turbulence and to determine the characteristic
structures seen during the relaminarisation process itself.
Currently, such calculations would be prohibitively expensive for
fully 3-dimensional models. Severe truncation to only a few Fourier
modes in the tilted cross-wise direction has also proven useful in
determining the origin of oblique bands in plane Couette flow
\citep{barkley07}.

The aim of this article is to establish a model system which
preserves the rich dynamics of pipe flow but reduces the number of
degrees of freedom of the system so considerably that the two
current issues mentioned above  can be probed. The price paid for
this reduction is, of course, a close quantitative match with  fully
3-dimensional pipe flow. But this is more than counterbalanced by
the ability to isolate and explore, for example, a `puff-like'
structure in a much more accessible system. The issues to be
addressed {\em within} the reduced system are as follows. ({\it a})
Do the relaminarisation statistics for turbulent puffs differ in
character between short and long pipes?
 In particular, do long-pipe simulations indicate a critical $\Rey$  for
 sustained puffs whereas short-pipe simulations not? And
({\it b}) what does the attracting `edge' state in the
laminar-turbulent boundary look like in a long pipe? Is it localised
like a puff and, if so, does it delocalise at the same $\Rey$ as a
puff?

Previous attempts to find such a reduced model have focused on
axisymmetric pipe flow \citep{patera81} and helical pipe flow
\citep{landman90,landman90b}, but in both cases the subcritical
dynamics of pipe flow is not retained. We briefly revisit these
calculations to search afresh for evidence of turbulent transients
before introducing a new 2$+\epsilon$-dimensional model which retains the
salient features of fully 3-dimensional pipe flow. The presentation
starts by discussing the formulation used for the calculations
performed throughout this and earlier work
\citep{willis07,willis08}.

\section{Formulation}

Given diameter $D$ and fixed mean axial speed $U$,
it is numerically convenient to scale lengths by $\frac{1}{2}D$ and
velocities by $2U$ in the Navier--Stokes equations, leading to
\begin{equation}
   \label{eq:govnonD}
   (\pd{t} + \vec{u}\cdot\grad) \vec{u}
   = -\grad p + \frac{4}{\Rey}\,(1+\beta)\,\vechat{z}
    + \frac{1}{\Rey}\bnabla^2 \vec{u} ,
\end{equation}
where the non-dimensional variable $\beta$ is the fractional
pressure gradient, additional to the laminar flow, required to
maintain a steady $U$. A Reynolds number $\Rey_p$, based on the
applied pressure gradient, is given by $\Rey_p = \Rey\,(1+\beta)$.
Our numerical formulation is based on the potential formulation of
\cite{marques90}, which is further re-expressed to ease numerical
solution.  An averaging operator is introduced in the axial
direction, $z$, which is periodic over a length $L=2\pi/\alpha$,
\begin{equation}
   P_z \,( \,\cdot\,) = \frac{1}{L}\int_0^L \,(\,\cdot\,)\,
   \, {\mathrm d}z .
\end{equation}
The velocity, $\vec{u}$, is then expressed in terms of a potential
$\psi=\psi(r,\theta,z)$,
the axially independent flow $h=h(r,\theta)$ and a
purely axially-dependent potential
${\phi}={\phi}(r,\theta,z)$,
\begin{equation}
      \vec{u} = h\,\vechat{z} + \curl(\vechat{z}\psi)
      + \curl\curl(\vechat{z}{\phi}) ,
\end{equation}
such that $P_z\,\phi=0$.
Writing the nonlinear terms as
$ 
   \vec{b} = (\vec{u}\cdot\grad)\,\vec{u} ,
$ 
the governing equations become
\begin{eqnarray}
   \label{eq:govPot}
   & (\pd{t} - \frac{1}{\Rey}\laplace)\, h
   \,=\, -P_z \, \vechat{z}\cdot\vec{b} , \nonumber \\
   & (\pd{t} - \frac{1}{\Rey}\laplace)\,\laplace\laplace_h{\phi}
   \,=\,  - (1 - P_z) \, \vechat{z}\cdot\curl\curl\vec{b} , \\
   &(\pd{t} - \frac{1}{\Rey}\laplace)\, \laplace_h \psi
   \,=\, \vechat{z}\cdot\curl\vec{b} ,
   \nonumber
\end{eqnarray}
where $\laplace_h := \laplace-\pd{zz}$.  For a boundary condition
$\vec{u}=\vec{g}(\theta,z)$, conditions on the potentials are
\begin{eqnarray}
   \label{eq:bcs}
   & h = 0, \quad \phi=0, \quad -\pd{r}\psi=g_\theta, \quad -\laplace_h\phi=g_z, \\
   & \frac{1}{r}\,\pd{\theta}\psi + \pd{rz}\phi = g_r,
   \quad
   \pd{rz}\laplace_h\psi
   -\frac{1}{r}\pd{\theta}\laplace\laplace_h\phi =
   \Rey\, \vechat{r}\cdot\curl\vec{b} - \pd{zz}\,\vechat{r}\cdot\curl\vec{g}
   \nonumber
\end{eqnarray}
(see appendix \ref{app:BCs} for details). Note that
$\vechat{r}\cdot\curl\vec{b}=0$ on the boundary unless an internal
body force is added.
Variables are expanded in Fourier modes,
\begin{equation}
   A(r,\theta,z)=
   \sum_{k,m}A_{km}(r)\,\exp(\mathrm{i}\alpha kz+\mathrm{i}m\theta).
\end{equation}
As the variables are real, their coefficients satisfy the property
$A_{km}=A^*_{-k,-m}$, where $^*$ indicates the complex conjugate,
and therefore only coefficients with $m\ge 0$ are kept. Numerical
truncation in $k$ and $m$ is discussed in following sections. The
operator $P_z$ picks out $k=0$ modes and $(1-P_z)$ retrieves $k\neq
0$ modes. In addition to the boundary conditions (\ref{eq:bcs}),
regularity at the axis imposes symmetries on the Fourier modes
across the axis.  For the potentials, each mode is even(odd) in $r$
if $m$ is even(odd).

The system for $h$ is simple to solve as it is second order and has
two boundary conditions on $h$, one at the boundary and a symmetry
condition at the axis.  The system for $\psi$ and $\phi$ is more
difficult to invert as it is coupled through the boundary condition.
To enable numerical solution we reformulate the system for $\psi$
and $\phi$ into a set of five equations, each second-order in $r$,
and use an influence matrix technique to by-pass the coupled
boundary condition (see appendix \ref{app:reformulated} for details).

Both finite-difference and Chebyshev expansions have been used in
radius.  The latter is better at low radial resolution, but the
former involves only banded matrices (a 9-point stencil is used),
requiring less memory, and is faster for high radial resolutions.
Time discretisation is second-order using Crank--Nicolson for the
diffusion term and an Euler predictor step for the non-linear terms.
Information from a Crank--Nicolson corrector step is used to control
the timestep size.  Nonlinear terms, $\vec{b}$, are evaluated using
the pseudo-spectral method and are dealiased using the
$\frac{3}{2}$-rule. The code was tested to reproduce eigenvalues
about the laminar state, eigenvalues about nonlinear travelling wave
solutions from \cite{wedin04}, by direct comparison with a primitive
variable code \citep{kerswell07} during the relaminarisation of a
perturbed travelling wave, and to calculate the turbulent statistics
of \cite{eggels94}.

\section{The absence of turbulence in previous models}

The original calculations for axisymmetric pipe flow
\citep{patera81} and for helical flow \citep{landman90} found no
evidence for turbulence or even long transients at $\Rey \leq 4000$.
Here we show that this
conclusion extends to $\Rey$ as large as $10^5$ and for huge initial
disturbances suggesting that there are no exact unstable solutions beyond
Hagen-Poiseuille flow within these dynamical subspaces.

The axisymmetric model is straightforward to simulate using the
numerical algorithm described  above by time stepping only modes
with $m=0$. In helical flow, variations in $\theta$ and $z$ are
reduced to a dependency in the one variable $\xi=\theta+\alpha z$,
where $\alpha$ is the pitch of the helix and periodicity over
$L=2\pi/\alpha$ is preserved. In this scenario the flow may be
expanded $\vec{u}(r,\xi)=\sum_q \vec{u}_q(r)\exp(\mathrm{i}q\xi)$,
corresponding to taking only modes $k=m\to q$ in our formulation.
Letting $\vec{u}'$ be the deviation from the laminar flow,
random initial disturbances of the form
\begin{equation}
   \vec{u}' = \sum_{k^2+m^2\neq 0} r^2(1-r)^2 \,
   (\alpha^2k^2+m^2)^{-\frac{1}{2}} \,
   \vec{a}_{km} \exp(\mathrm{i}\alpha kz+\mathrm{i}m\theta),
\end{equation}
were applied to the flow (after projection onto the space of
solenoidal functions to enforce incompressibility), where the
components of $\vec{a}_{km}$ were random numbers in $\mathbb{C}$
s.t.\ $|a_{km}|=1$.

\begin{figure}
   \psfrag{t}{$t\,(D/U)$}
   \psfrag{E3D/E0}{$E'_{k\ne 0}/E_0$}
   \epsfig{figure=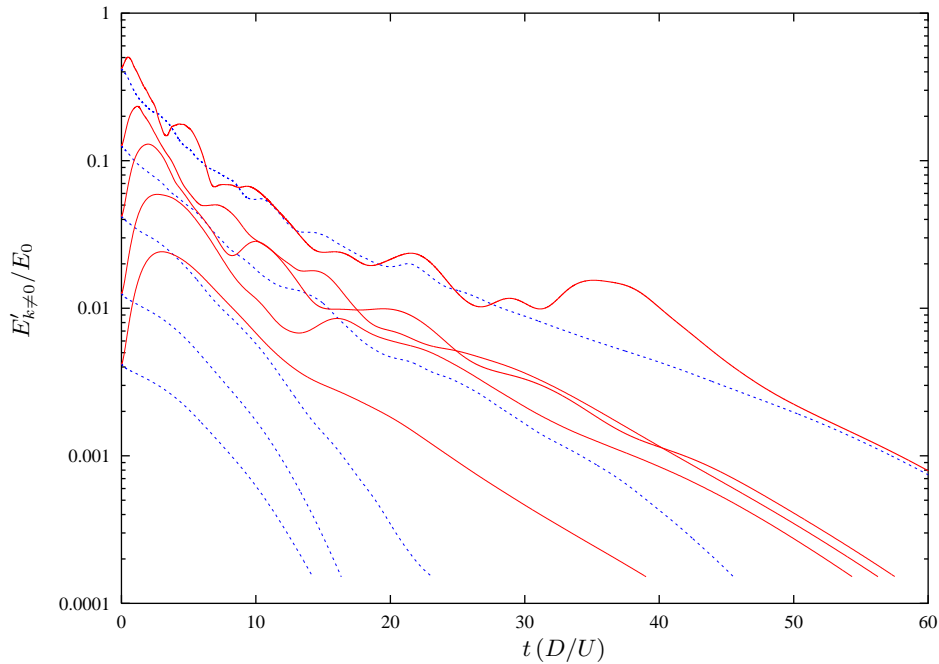}
   \caption{\label{fig:helaxi}
      Decay of axisymmetric (dashed) and helical (solid) perturbations, initially
      of energy up to 40\% of the laminar profile ($\Rey=10,000$
      and pipe $2 \pi \, D$ long).
   }
\end{figure}
Figure \ref{fig:helaxi} shows time evolution of random disturbances
at $\Rey=10\,000$ for a pipe of length $2\pi$ diameters. Initial
disturbances were normalised for $E'_{k\neq 0}/E_0$ up to $0.4$,
where $E'_{k\neq 0}$ is the energy of the axially-dependent modes
and $E_0$ is
the energy of the laminar profile; 
i.e.\ disturbances of up to 40\% of the energy of the laminar flow
were considered.  By way of comparison, the turbulent test case of
\cite{eggels94} has only $E'_{k\neq 0}/E_0 \approx 0.014$ at
$\Rey=5300$. The number of finite difference points in radius and
truncation of the Fourier modes was $(160,\pm 256)$ in $(r,z)$ and
$(r,\xi)$; the large radial resolution was required to stably solve
for such high initial energies. Axisymmetric disturbances show
little sign of nonlinear interactions at this $\Rey$ and decay
almost monotonically. Five other sets of runs for other random
disturbances showed similar behaviour. Helical flow is slightly more
promising showing occasional moments of growth against a dominant
decay.

\begin{figure}
   \psfrag{t}{$t\,(D/U)$}
   \psfrag{E3D/E0}{$E'_{k\ne 0}/E_0$}
   \epsfig{figure=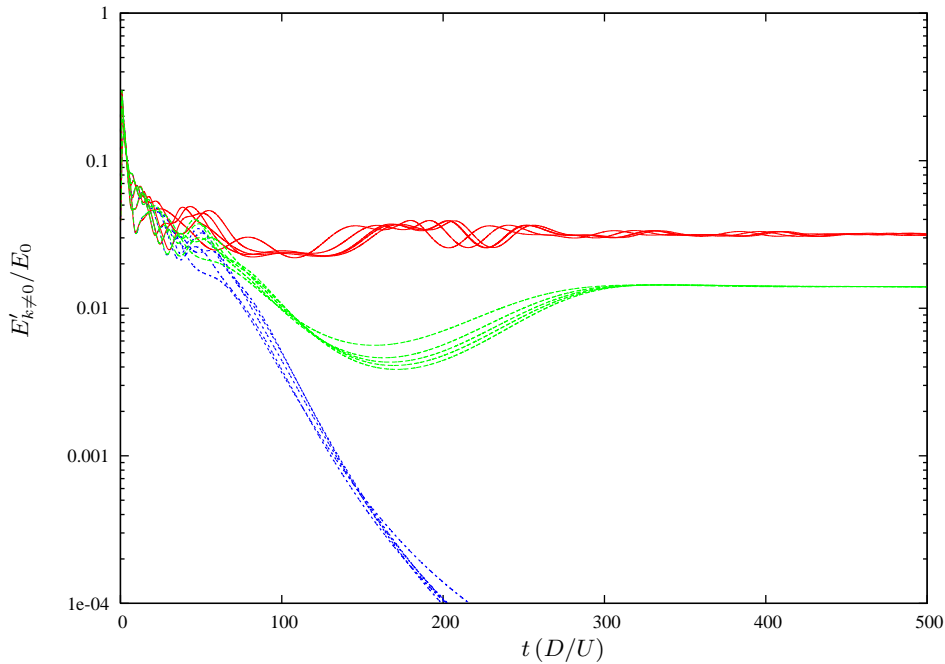}
   \caption{\label{fig:rothel}
      Large helical disturbances scaled from 4\% to 20\% of $E_0$
      rapidly settle to parabolic flow
      for $R_\Omega=-50$ and to
      laminar helical waves for $R_\Omega=-100$, $-300$ ($\Rey=4000$, $L=16\pi\,D$).
   }
\end{figure}

Helical flow exhibits a linear instability at relatively small
rotation rates about the axis, which occurs at longer $L$ for larger
$\Rey$ \citep{mackrodt76}. This supercritical bifurcation and the
subsequent nonlinear waves \citep{toplosky88,landman90b} provided an
excellent test of the helical code. Measuring the rotation rate by
$R_\Omega = \frac{1}{4}\,\Omega\,\Rey$, where the angular velocity
$\Omega$ is in units $U/D$, there is a supercritical bifurcation to
helical waves at $R_\Omega=-52.43$, $\Rey=4000$ and $L=16\pi\,D$.
Figure \ref{fig:rothel} shows evolution of five random initial
disturbances of $E'_{k\neq 0}/E_0$ up to $0.3$ in the presence of
rotation; truncation at $(80,\pm 128)$ in $(r,\xi)$. For
$R_\Omega=-50$ the flow quickly returns the the parabolic profile.
At $R_\Omega=-100$ and $-300$, well beyond the linear instability, the
flow rapidly returns to a finite-amplitude helical wave flow. This
supercritical behaviour persists for modest rotations so that no
long term turbulent transients can be generated. \cite{barnes00}
have shown that these helical waves themselves undergo a
supercritical Hopf bifurcation so that solutions cannot obviously be
traced back to non-rotating flow. Figure \ref{fig:rothel} suggests
that disconnected branches which could lead to subcritical
turbulence in  rotating helical pipe flow are unlikely to exist.

Our calculations suggest that rotating helical pipe flow follows the
supercritical route to turbulence via a sequence of supercritical
bifurcations, rather than the abrupt subcritical behaviour of
3-dimensional non-rotating pipe flow. Having seen strong decay at
$\Rey$ approximately five times that for which turbulence is
observed in the laboratory, it appears that neither dynamics
restricted to helical or axisymmetric subspaces are relevant for the
observed transition.

\section{A 2$+\epsilon$-dimensional model}

We now introduce a third model which has high resolution in the
cross-stream (radial) and streamwise (axial) directions, but only a
few modes in the spanwise (azimuth) direction. The model was chosen
to preserve a high radial resolution as streak features close to the
wall appear to be important in the self-sustaining process as do
detachments from the wall during the relaminarisation stages of
low Reynolds number turbulence. High axial resolution was retained
to allow the possibility of localised turbulent structures. This
left only the azimuthal direction in which to reduce the number
of degrees of freedom: only Fourier
modes $m=0,\pm m_0$ were considered, which corresponds to a
sinusoidal variation in azimuth, an azimuthal shift of the sinusoid
and a mean mode.

\begin{figure}
   \begin{center}
      \epsfig{figure=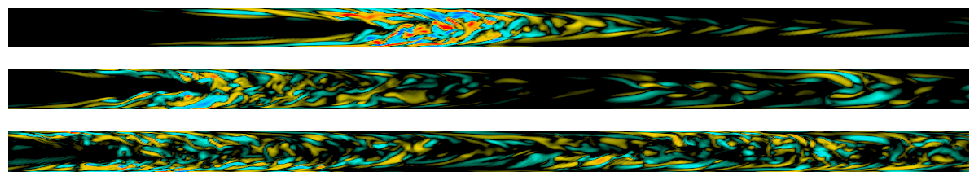, angle=90, width=36.66mm}
      \hspace{3mm}
      \epsfig{figure=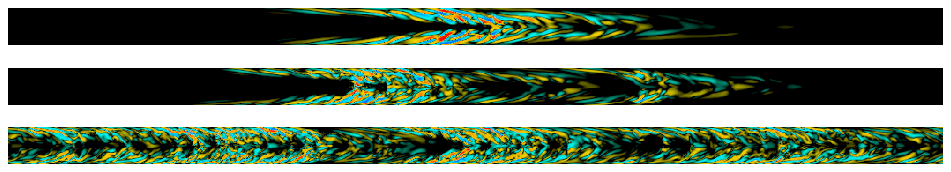, angle=90, width=37.0mm}
      \hspace{2mm}
      \psfrag{z(D)}{$z\,(D)$}
      \psfrag{10Ert}{$10\,E(u_r,u_\theta)$}
      \psfrag{Ez}{$E(u_z)$}
      \epsfig{figure=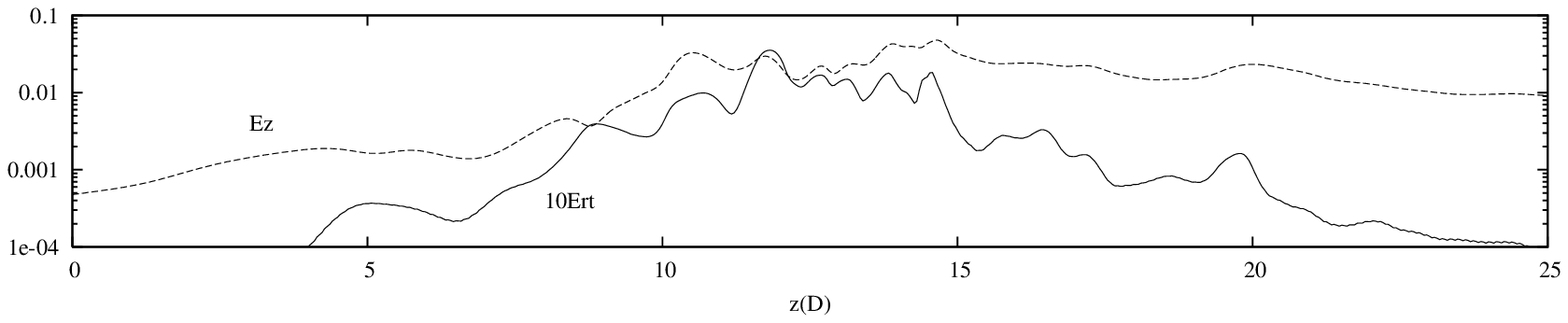, angle=90, width=38mm}
   \end{center}
   \caption{\label{fig:puffslug}
      Axial vorticity in the $(r,z)$-plane, 1:1 aspect ratio,
      flow up the page; 
      only $25\,D$ of a periodic $50\,D$ domain shown.
      {\it From left to right}:
      fully 3-dimensional simulations at $\Rey=2000$, $2300$, $2700$ and
      2$+\epsilon$-dimensional simulations at $\Rey=2600$, $3200$, $4000$.
      Presented for both models are localised `puffs', the early stages of
      delocalisation by the generation of a second puff downstream,
      and global slug-turbulence at larger $\Rey$.
      {\it Far right}:
      Energy in separate components of the velocity as a function of axial
      position (units $D^2U^2$) for the model puff snapshot at $Re=2600$.
   }
\end{figure}

\subsection{Spatial characteristics}

Localised structures, surprisingly similar to puffs, were captured
by the 2$+\epsilon$-dimensional model. Figure \ref{fig:puffslug}
compares a puff structure from a simulation fully resolved in
azimuth (all $m$ up to $\pm 24$) with a `puff' from the
2$+\epsilon$-dimensional model ($m=-3,0,3$). The plots are of the
correct aspect ratio but only half of the computational domain is
shown.  Puffs from the model appear to be similar to resolved puffs
in both length and structure, having a smooth upstream region close
to the wall, an active turbulent region, and a dissipative region
downstream (although note the different $Re$). Only modest radial
resolutions were required to observe such structures: a spectral
resolution of 35 Chebyshev modes was used for the calculations of
this section. Several calculations were performed with a lower
resolution of 25 radial modes, but puff structures tended to
elongate, requiring a longer pipe and thus offsetting 
 the reduction in computation times.  The energy plot in
figure \ref{fig:puffslug} shows that the while the axial
deviation from the mean flow is extended, $20\,D$ or greater,
the roll components are highly localised, extending only $5$-$10\,D$.
Such localisation of the active region of the flow has been
observed in the full 3-dimensional case \citep{willis08}.

Axial resolution was chosen to approximately match the spectral
drop-off in $r$ (approximately 4 orders in the magnitude of the
spectral coefficients, or 8 orders in the power spectrum) and was
$\pm 384$ for $L=16\pi\,D \approx 50$ diameters. Axial resolution was
changed proportionally for other $L$ considered in the rest of this
section, hence keeping the smallest resolved scale fixed. Puffs were
found to translate within 2\% of $U$, slightly faster than in
3 dimensions, where they travel approximately 10\% slower. Also shown in figure
\ref{fig:puffslug} is that the transition from localised to global
turbulence is gradual, as observed experimentally. At larger $\Rey$
the puff becomes delocalised, splitting into two or more localised
turbulent regions with relatively laminar regions in between. At
much larger $\Rey$ the proportion of vigorous turbulence is seen to
increase, as recorded by \cite{gilbrech65}.

\subsection{Temporal characteristics}
\label{sect:temporal}
\begin{figure}
   \psfrag{t}{$t\,(D/U)$}
   \psfrag{E3D/E0}{$E'_{k\ne 0}/E_0$}
   \epsfig{figure=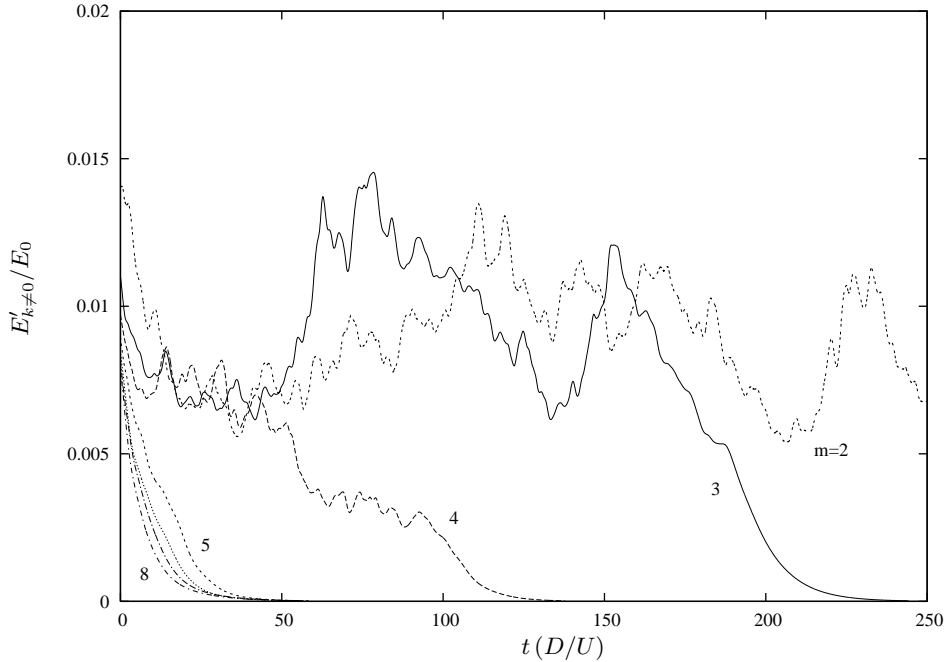}
   \caption{\label{fig:otherm}
      Initial trials of the 2$+\epsilon$-dimensional model for several $m$-fold rotational
      symmetries.
   }
\end{figure}
Another important feature captured by the model is that localised
puffs may survive for long times before a sudden decay as observed
experimentally \citep{peixinho06}. Typical transients for different
$m_0$ are shown in figure \ref{fig:otherm} which indicates that
higher rotational symmetries tend to decay more quickly.  As
structures of 3-fold rotational symmetry are the most frequently
observed for transitional $\Rey$
\citep{duggleby07b,schneider07,kerswell07,willis08}, $m_0=3$ was
chosen for analysis of the lifetimes of disturbances. For this
$m_0$, azimuthal length scales are also comparable to the radial
length scale in the model.
\begin{figure}
   \psfrag{T}{$T\, (D/U)$}
   \epsfig{figure=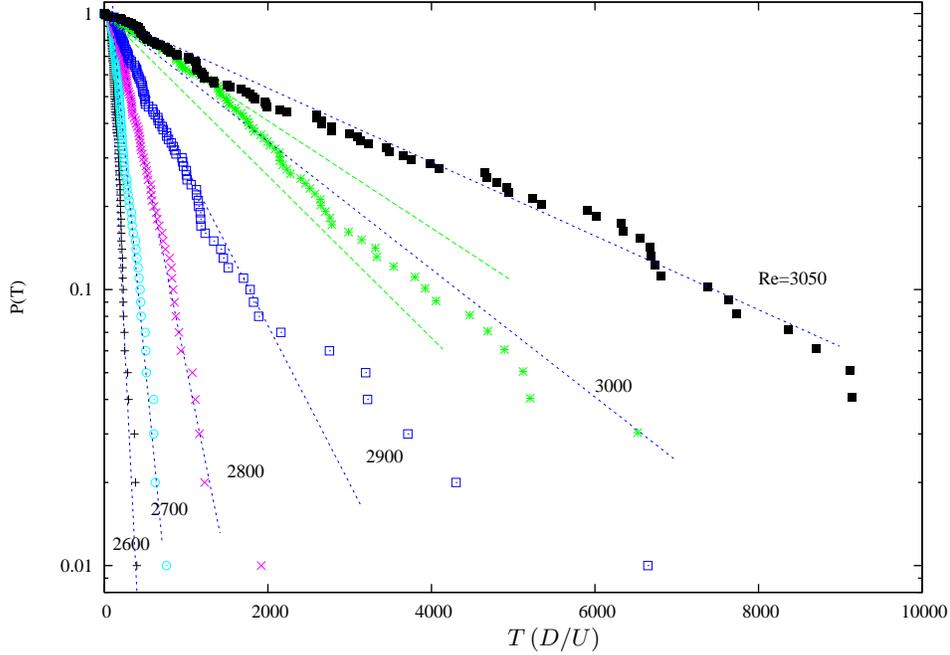}
   \caption{\label{fig:PT32piD}
      Probability of transient surviving to time $T$
      in the $m_0=3$ model for a pipe of $\approx100$ diameters ($L=32\pi\,D$);
      100 observations per $\Rey$. See the text for how the
      `best fit' lines were drawn.
   }
\end{figure}
Figure \ref{fig:PT32piD} shows the probability distribution function
for puff lifetimes based upon 100 runs at each of several $\Rey$ in
a pipe $\approx 100$ diameters long ($L=32\pi\,D$). Sets of initial
puff conditions were generated from snapshots of a long run at a
sufficiently large $\Rey$, similar to the annealing procedure
adopted in \cite{peixinho06} and \cite{willis07}. No dependence on
the initial condition was observed, however, other than in the very
early times of the transient. For the model, times of order $10^4\,
D/U$ could be achieved, significantly longer than achieved in
\cite{willis07} and in less computation time. The log-plot shows an
exponential distribution indicating a memoryless process.

Slopes in figure \ref{fig:PT32piD} are based on a median time,
$\tau$, calculated by the bootstrapping method adopted in
\cite{willis07com}.  
The shorter slopes represent 95\% confidence
intervals for the slope of the data at $\Rey=3000$, and 
are accurately calculated by the bootstrapping method.
The best-fit line was found to be poor estimator of the mean, 
being overly sensitive to rare events
(e.g. the outlier in figure \ref{fig:PT32piD} for $\Rey=2900$), 
and the goodness-of-fit does not provide the correct error estimate.
In the bootstrapping method,
samples of size $N$ are generated by resampling from the original
$N$ observations, with equal probability of selecting each. This is
repeated $100\,000$ times and the distribution of the means of the
samples provides an accurate confidence interval for the mean of the
original data set.  When all data fall within a maximum observation
time, the confidence intervals generated by this method converge to
those predicted by the central limit theorem. A 95\% confidence
interval for $\tau$ is approximately $\tau \pm 2 \tau/\sqrt{N}$. 
The method is
particularly useful when the data is truncated in time and therefore
a mean cannot be directly calculated. The bootstrapping method
easily accommodates such data by further resampling when a truncated
point is chosen. Any additional error associated with the extra
resampling is reflected in a wider confidence interval.

Estimating the median lifetime of a puff from the data is subject to 
two sources of error: the initial transition period during which
the flow evolves from the initial condition to become a puff,
and the presence of a final relaminarisation phase. The former
is eliminated by considering each of the first observations as an 
initial cut-off time and by examination of the effect of the cut-off 
on the estimator $\tau$. See \cite{willis07com} for an example plot 
of $\tau$ with confidence intervals {\it vs.}\ number of observations cut. 
Removing the first few observations eliminates the effect of the
transient on $\tau$ but slightly widens the confidence interval.  
The relaminarisation time error was minimised by 
identifying a threshold 3D energy below which the turbulent flow 
always relaminarises and applying the same value to all runs to 
indicate the end of the puff lifetime 
($E'_{k\neq 0}/E_0 = 0.001$).

The median time, $\tau\,(D/U)$, is dependent on the parameter $\Rey$.
As mentioned above, whether $\tau$
diverges to infinity or not at a finite $\Rey$ is a matter of
ongoing debate. In laboratory experiments using a pipe with
$D=20\mathrm{mm}$, $L=785\,D$, \cite{peixinho06} found evidence that
$\tau\sim 1/(\Rey_c-\Rey)$ with $\Rey_c=1750$. 
In contrast, experiments by \cite{hof06}
for $D=4\mathrm{mm}$, $L=7500\,D$, and using a different method to
initiate the puff, found that $\tau\sim \exp(c_1\Rey)$, for some
constant $c_1$. Numerical experiments by \cite{willis07,willis07com}
using well-resolved puffs in a $50\,D$ periodic pipe, however, show
lifetimes to be significantly different from the exponential scaling
and the simple power $-1$ was clearly seen with $\Rey_c=1870$,
overestimating the experimental value of 1750 \cite{peixinho06} by
only 7\%. Computational and experimental limitations have confined
observations of $\tau$ to $O(10^3-10^4) \,D/U$.

\begin{figure}
   \psfrag{1/t}{$1/\tau$}
   \psfrag{6D}{$L=2\pi\,(D)$}
   \psfrag{13D}{$4\pi$}
   \psfrag{25D}{$8\pi$}
   \psfrag{50D}{$16\pi$}
   \psfrag{100D}{$32\pi$}
   {\it a}\epsfig{figure=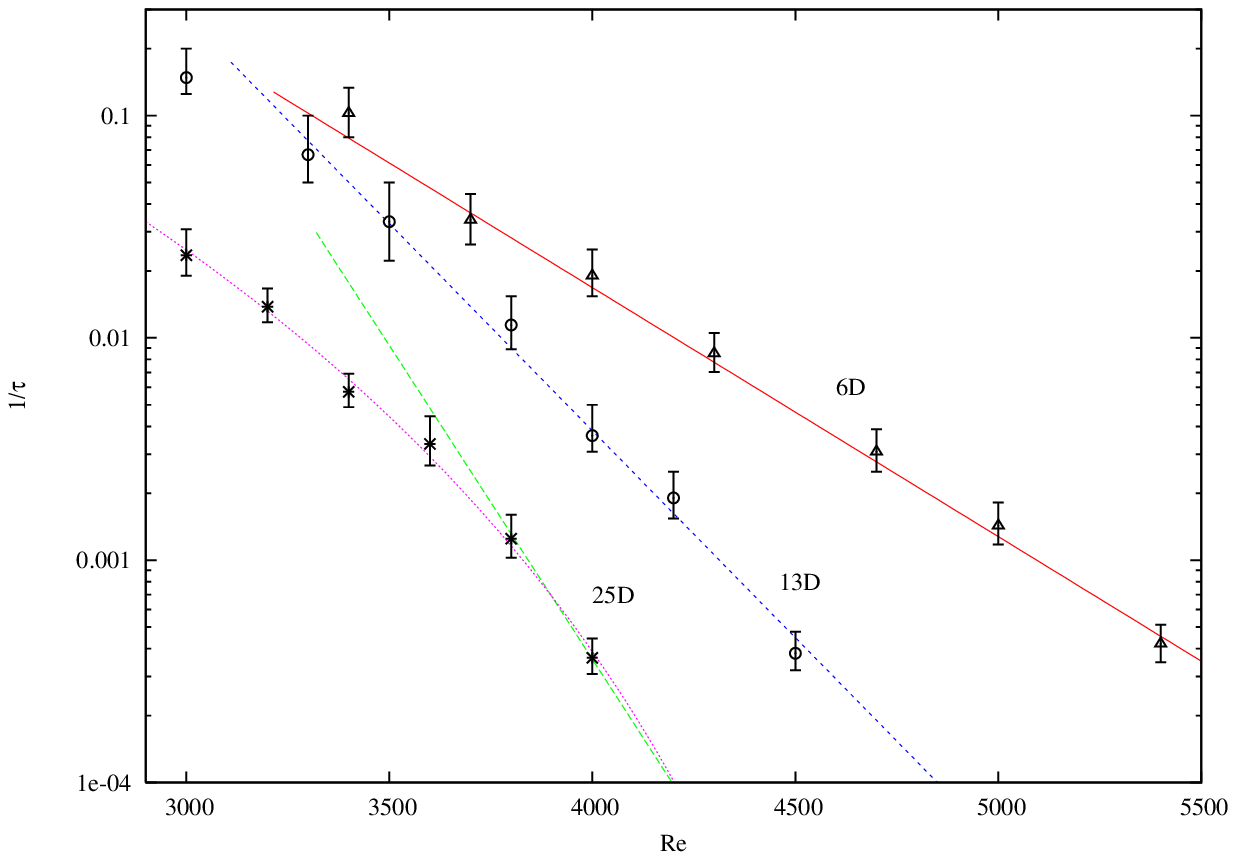}\\
   {\it b}\epsfig{figure=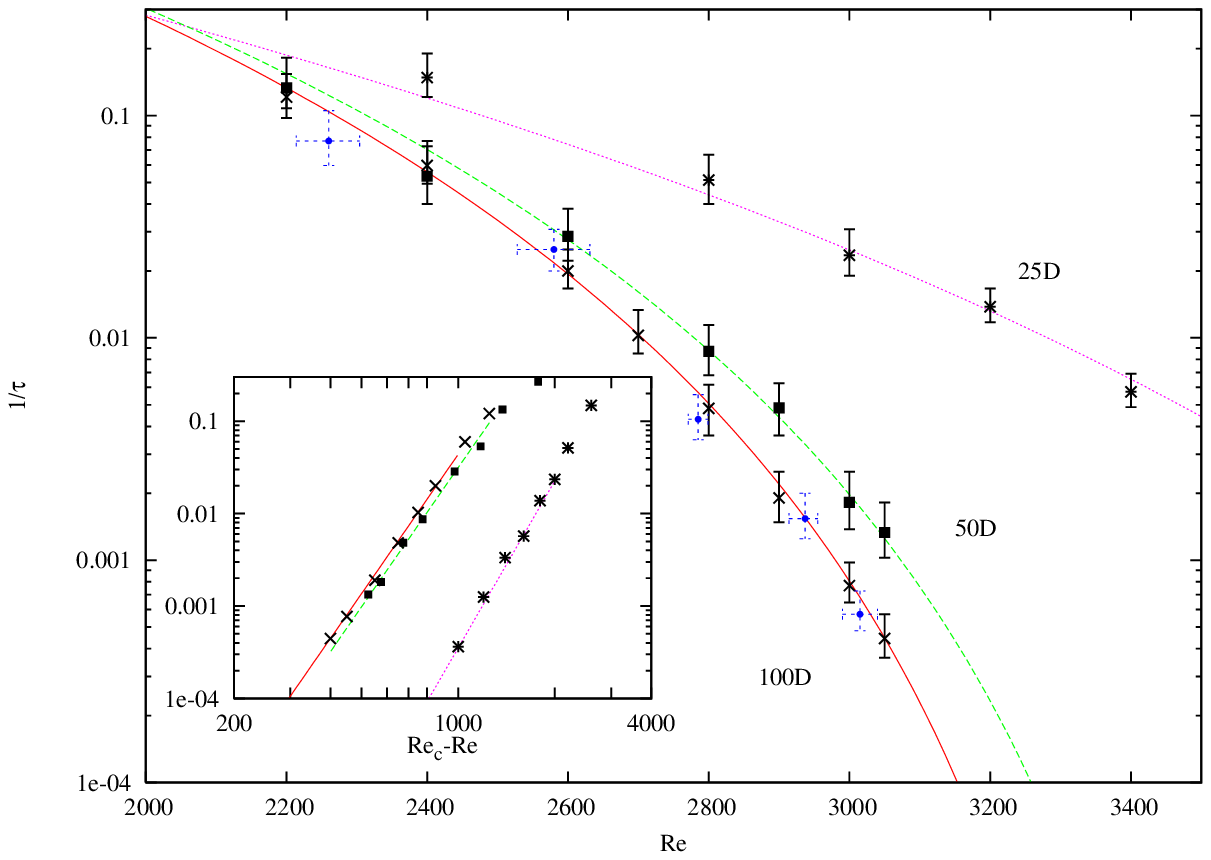}\\
   \caption{\label{fig:ReTmed1log}
      Sensitivity of the lifetime, $\tau\,(D/U)$, of transients to the pipe length.
      The upper plot ({\it a}) shows the data for pipe lengths $2 \pi$, $4
      \pi$ and $8 \pi$ (in $D$) and exponential fits, $1/\tau = \exp(a \Rey+b)$,
      through all points which
      correspond to global (slug) turbulence. The 4 leftmost data points
      for $8 \pi$ correspond to localised (puff) turbulence and
      a $1/\tau \sim (\Rey_c-\Rey)^{\beta}$ fit is shown for these.
      The lower plot ({\it b}) shows data where localised (puff) turbulence is
      present for pipe lengths $8 \pi$, $16 \pi$ and $32 \pi
      \approx 100$ diameters long.  Here the best fit lines take the form
      $1/\tau \sim (\Rey_c-\Rey)^{\beta}$ where the
      $1/\tau$ vs $(\Rey_c-\Rey)$ plot on log-log scales ({\it inset})
      shows these fits as straight lines (the same best fit line for
      $L=8 \pi\,D$ is plotted in both ({\it a}) and ({\it b})).
      Points with error bars in $Re$ were calculated using a fixed pressure
      gradient as opposed to fixed mass flux. This data is plotted
      using the time-averaged $\Rey$
      for a $32 \pi$ $D$ pipe and falls precisely on the fixed mass-flux data
      (as for all other points, 100 observations were used for each).
   }
\end{figure}
\begin{table}
   \begin{center}
   \begin{tabular}{cccccr}
       & \multicolumn{2}{c}{Global Turbulence}
     & \hspace{0.5cm} & \multicolumn{2}{c}{Localised Turbulence} \\
       & & &&& \\
   $L\,(D)$ & $a$ & $b$            &     & $\beta$ & $\Rey_c$ \\
            &     &                &     &         &          \\
   $2\pi$ & $-2.5\times10^{-3}$ & $6.2$ & \\
   $4\pi$ & $-4.3\times10^{-3}$ & $11.6$ & \\
   $8\pi$ & $-6.5\times10^{-3}$ & $18.1$ && $5.9 \pm 2.0$ & $\approx 5000$\\
   $16\pi$ & &                           && $4.6 \pm 1.6$ & $3575$\\
   $32\pi$ & &                           && $4.8 \pm 1.6$ & $3450$
   \end{tabular}
   \end{center}
   \caption{\label{tab:expfits}
      Best fit parameters for an exponential fit $1/\tau=\exp(a\Rey+b)$
      and an algebraic fit
      $1/\tau =(\Rey_c-\Rey)^{\beta}$ for different
      lengths of periodic pipe.}
\end{table}

Figure \ref{fig:ReTmed1log} shows the maximum likelihood estimator
for $1/\tau$ and 95\% confidence intervals across a spectrum of pipe
lengths. Lifetimes an order of magnitude larger than the
calculations of \cite{willis07} were possible in a pipe twice as
long, $L\approx 100$ diameters.  In short pipes (figure
\ref{fig:ReTmed1log}{\it a}), where turbulence fills the domain and
is therefore global, the lifetime appears to follow an exponential
scaling $1/\tau=\exp(a\Rey+b)$ with values for $a$ and $b$ are given
in Table \ref{tab:expfits}.  The $8\pi\,D$ ($\approx 25$ diameter)
pipe represents a cross-over situation in that the turbulence is
only truly pipe-filling for the highest two data points whereas it
is localised for the four lower $\Rey$ shown in figure
\ref{fig:ReTmed1log}({\it a}). An exponential fit through just the
two higher $\Rey$ points, however, seems to fit a steepening trend
as the pipe lengthens suggested by the $2 \pi\,D$ and $4 \pi\,D$
data sets.

The remaining localised-turbulence points for $L=8\pi\,D$ are better
fit by an algebraic expression $\tau = 1/(\Rey_c-\Rey)^\beta$ (see
Table 1 for $\Rey_c$ and $\beta$).  The fit includes two extra
points at lower $\Rey$ for the $8 \pi\,D$ pipe shown in figure
\ref{fig:ReTmed1log}({\it b}). Doubling to a $16 \pi\,D$ ($\approx
50$ diameter) pipe produces a rapid drop in $1/\tau$ and a more
clearly defined curvature in the data. This length increase is
significant because the turbulent `puffs' in this reduced model are
approximately $20\,D$ long (see figure \ref{fig:puffslug}). Hence,
while such puffs will be significantly affected by enforced
periodicity over $\approx 25$ diameters, this artificial constraint
should be substantially relaxed in an  $\approx 50$ diameter pipe
and almost absent in a pipe of $\approx 100$ diameters. By way of
confirmation, a $32 \pi\,D$ pipe produces very similar
relaminarisation data to the $16 \pi$ $D$ pipe with both being
fitted well by the relation $\tau\sim 1/(\Rey_c-\Rey)^\beta$ based
upon similar values for the fitting parameters (see Table 1). Even
though the exponent $\beta$ is relatively poorly constrained, it is
clearly different from the value of 1 which is observed in fully
3-dimensional simulations \cite{willis07} and experiments
\cite{peixinho06}. This quantitative discrepancy is undoubtedly an
artifact of the reduced model and, in fact, is typical of other
models for different flows \citep{bottin98,lagha07}. What this
reduced model does clearly exhibit, however, is a {\em qualitative}
change in relaminarisation behaviour when the preferred localised
turbulent state (a `puff') is allowed to develop. If the pipe is
long enough to accommodate a `puff', the presence of a critical
$\Rey$ is suggested whereas when the pipe is shorter than  a `puff'
so that the turbulence is always global, no such critical threshold
is suggested.  This behavioural change is entirely consistent with
the seemingly contradictory results obtained recently using fully
3-dimensional Navier-Stokes calculations in short $5\, D$ pipes
\citep{hof06} and long $50 \, D$ pipes \citep{willis07}.

\subsubsection{Constant pressure gradient versus constant mass flux}

The reduced model was also used to investigate another open issue:
is the form of driving, being either a constant imposed
pressure-difference between the ends of the pipe
or a constant imposed mass-flux, important for the relaminarisation
behaviour of turbulent puffs? \cite{hof06} use a
pressure-drop set-up (constant mean pressure gradient)
in their experiments over a very long pipe whereas \cite{peixinho06}
suck fluid through
their pipe to produce a constant mass flux.
In an averaged sense, the two methods of driving the flow are equivalent
but instantaneously and locally the flow dynamics are different ---
the mass flux rate can fluctuate for a constant mean pressure-gradient flow, and
local variations in the pressure gradient cause fluctuations in the
total pressure-drop for a constant flux flow.
Hence puffs evolving in the two situations are subtly different,
and it is therefore a leading question as to whether they possess the same
relaminarisation behaviour.

To answer this, a series of constant pressure-gradient runs were
performed in the $32 \pi\,D$ pipe using randomly selected initial
conditions from the same long puff run as for the constant mass flux
runs. Again 100 runs were calculated for each data point at a given
pressure gradient and then the median lifetime plotted as a function
of $\Rey$ in figure \ref{fig:ReTmed1log}({\it b}). Note that a
horizontal error bar is plotted which indicates $\pm$ 
one standard deviation in the mean $\Rey$
value.  At the largest mean pressure-gradient used, 
the mean flow rate was $\Rey=3050$ with
fluctuations having one standard deviation of $\pm 25$. This error
in $\Rey$ should inversely scale with the
length of pipe for flow driven by a pressure head, provided that the
disturbance remains localised. For the long pipe of \cite{hof06}
such variations should be insignificant (1 part in $10^4$)
whereas for a short (numerical) periodic pipe very large
variations of O(10\%) can be expected. 
The new data points sit precisely on the $32 \pi\,D$ constant 
mass-flux curve indicating
that, for at least the long pipe ($\approx 100$ diameters), the precise
form of driving is unimportant for the probability of puff
relaminarisation.

\subsection{Characteristics of the laminar--turbulent boundary: short pipe}

Given an initially laminar flow, small perturbations decay back to
the laminar flow and larger perturbations develop into turbulence
for sufficiently large $\Rey$. This naturally leads to the question
of what characterises the dividing set of flows, for which a small
perturbation may lead to either laminar or turbulent flow.
\cite{itano01} used a shooting method to find such a boundary in
channel flow and discovered that the flow trajectory on the boundary
settled upon what they thought was a travelling wave solution, but
which was later identified as a slowly varying part of a periodic 
orbit \citep{toh03}.  
This orbit is stable within the
manifold of flows on this
laminar-turbulent boundary or `edge'. A similar situation is found
in plane Couette flow \citep{viswanath07,schneider08},
where a single simple
attractor is found. Pipe flow exhibits different characteristics,
however, with \cite{schneider07b} finding a chaotic attractor in
which trajectories pass nearby to exact travelling wave solutions
\citep{mellibovsky07,duguet07}. When the laminar-turbulent boundary
dynamics is restricted within certain symmetry subspaces, however,
simple attractors do emerge \citep{duguet07}.

Pipe flow is not so different from channel flow and yet they display
different dynamics on the boundary.  It is not difficult to imagine
therefore, that the severe truncation of our model could also lead
to a loss of chaotic behaviour on the boundary.  In this section
we show that the model preserves the chaotic end state for trajectories
on the boundary and that exact solutions exist.  This motivates
extension of the results to long computational domains, where undirected
3D calculations would be prohibitively expensive.

\begin{figure}
   \psfrag{Re=5000}{$\Rey=5000$}
   \psfrag{Re=6500}{$\Rey=6500$}
   \psfrag{Re=10000}{$\Rey=10\,000$}
   \psfrag{Re=16000}{$\Rey=16\,000$}
   \psfrag{b}{$\beta$}
   \psfrag{t}{$t\,(D/U)$}
   \psfrag{Ek0/E0}{$E'_{k\ne 0}/E_0$}
   \psfrag{E+}{$E^+$}
   \psfrag{E-}{$E^-$}
   \begin{tabular}{cc}
      \epsfig{figure=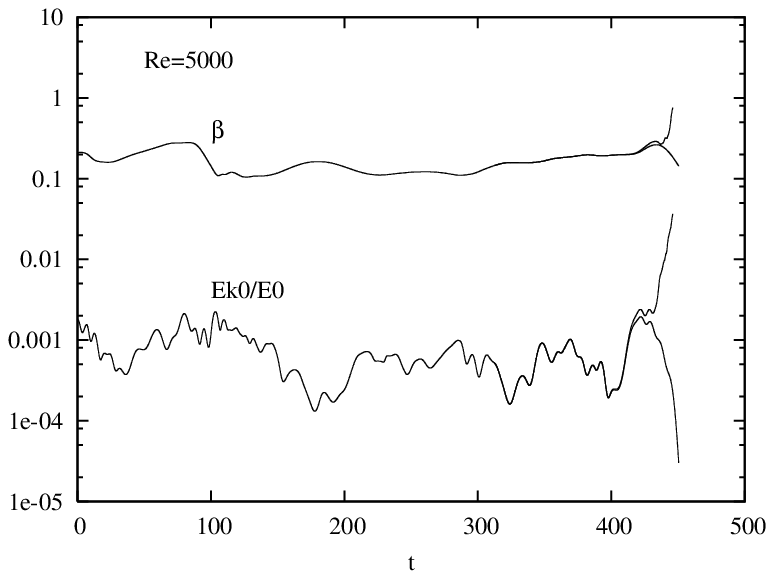, scale=0.83} &
      \epsfig{figure=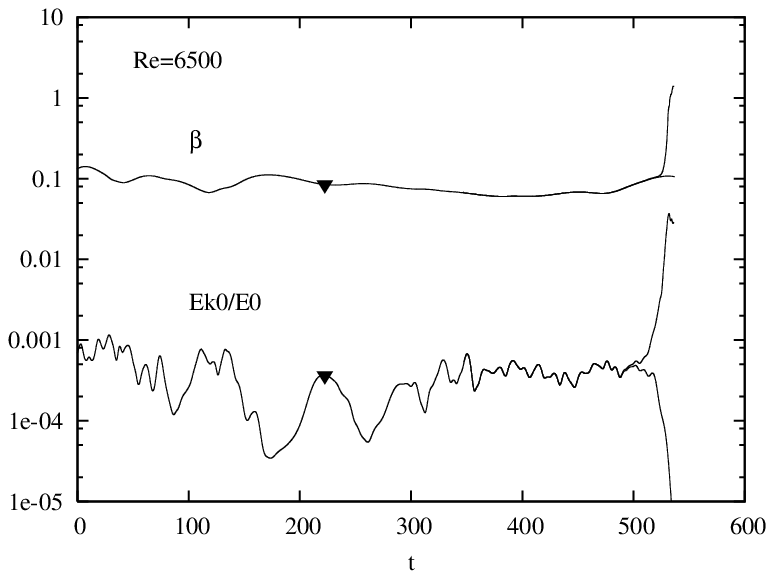, scale=0.83} \\
      \epsfig{figure=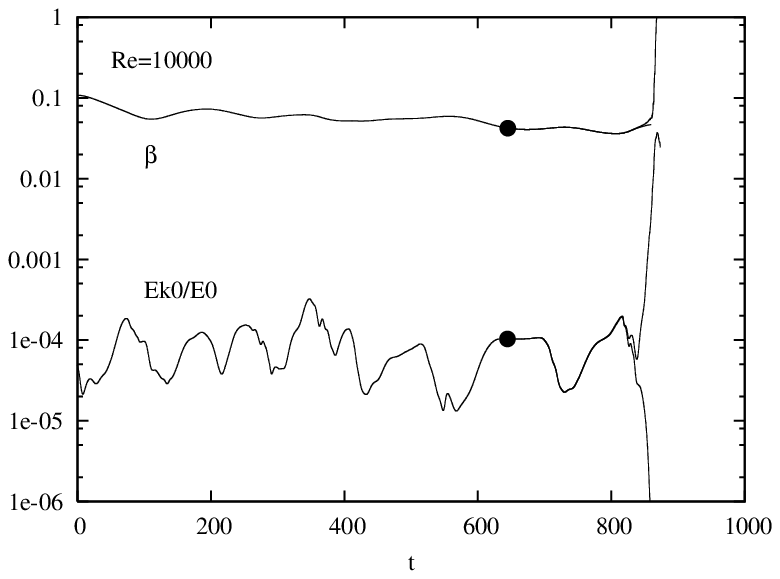, scale=0.83} &  
      \epsfig{figure=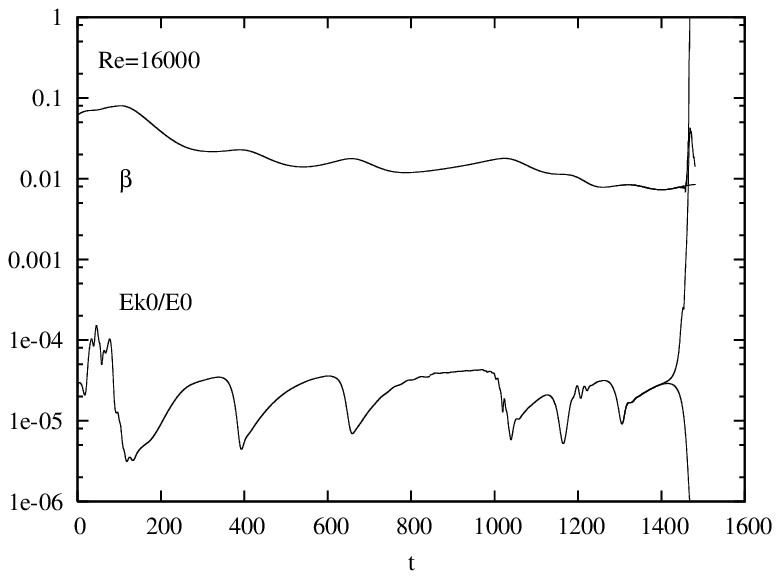, scale=0.83}
   \end{tabular}
   \caption{\label{fig:alp1edge}
      Edge trajectories for $L=\pi\,D$.  Exact travelling wave
      solution found nearby to the triangle at $\Rey=6500$
      (shown in figure \ref{fig:slices})
      has very similar structure to that found at the circle for
      $\Rey=10000$.
   }
\end{figure}

Boundary or edge trajectories for the model are shown in figure
\ref{fig:alp1edge}, for $L=\pi\,D$ and over a range of $\Rey$. While a
difference between the edge and developed turbulence may be seen by
a rapid increase in energy, a clearer measure appears to be $\beta$
related to the pressure gradient required to maintain the fixed flux (see
equation \ref{eq:govnonD}), or equivalently the friction.  The
pressure gradient is less than 10\% greater than the laminar value when
on the edge and is also smooth in time, whereas after a sudden
increase to turbulence it is as rapidly varying as the energy.

\begin{figure}
   \begin{tabular}{c}
     ({\it a}) \\
      \psfrag{-1}{}
      \psfrag{-0.8}{}
      \psfrag{-0.6}{}
      \psfrag{-0.4}{}
      \psfrag{-0.2}{}
      \psfrag{0}{}
      \psfrag{0.2}{}
      \psfrag{0.4}{}
      \psfrag{0.6}{}
      \psfrag{0.8}{}
      \psfrag{1}{}
      \epsfig{figure=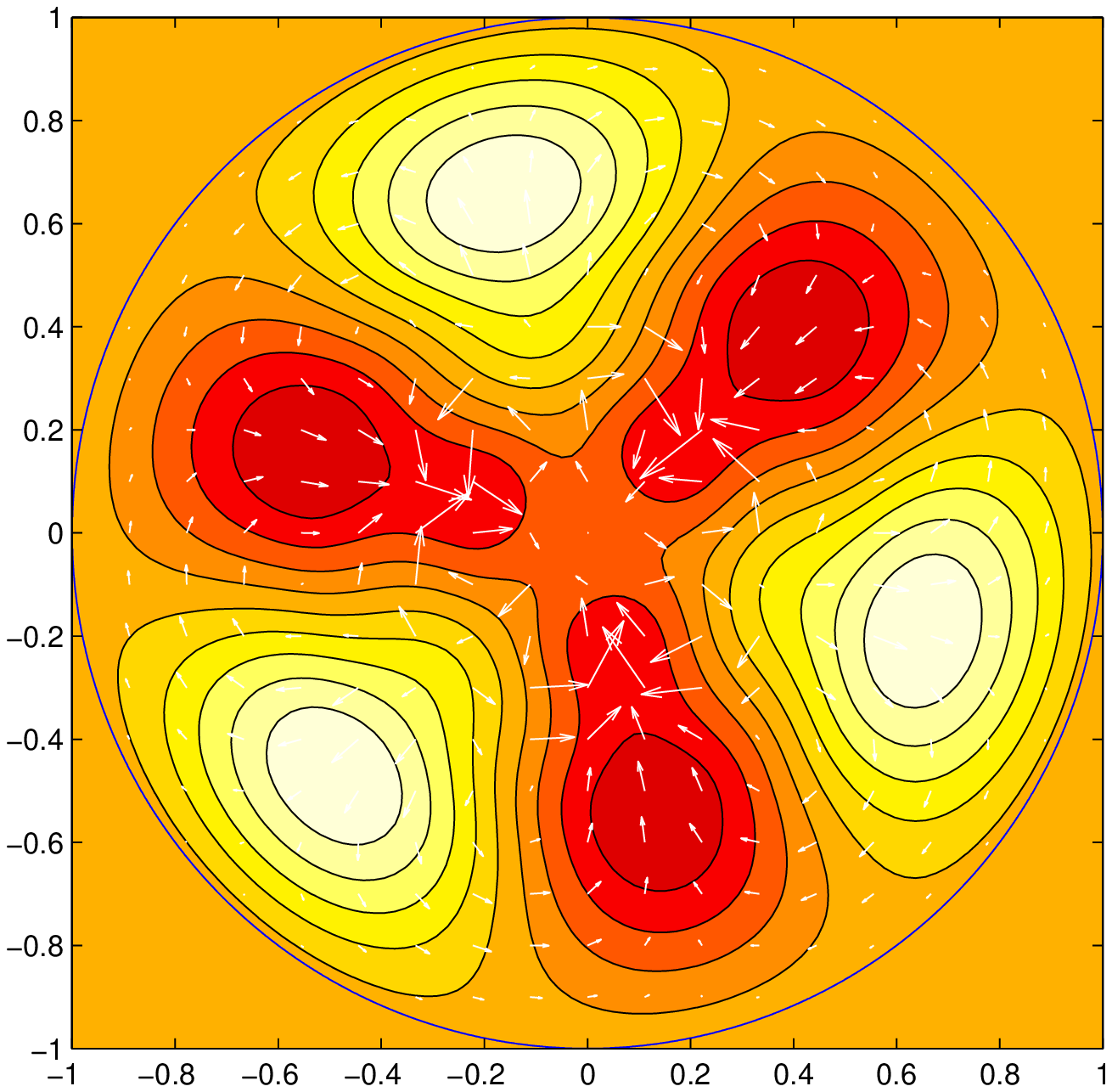, height=32mm}\\
      \psfrag{-1}{}
      \psfrag{-0.8}{}
      \psfrag{-0.6}{}
      \psfrag{-0.4}{}
      \psfrag{-0.2}{}
      \psfrag{0}{}
      \psfrag{0.2}{}
      \psfrag{0.4}{}
      \psfrag{0.6}{}
      \psfrag{0.8}{}
      \psfrag{1}{}
      \epsfig{figure=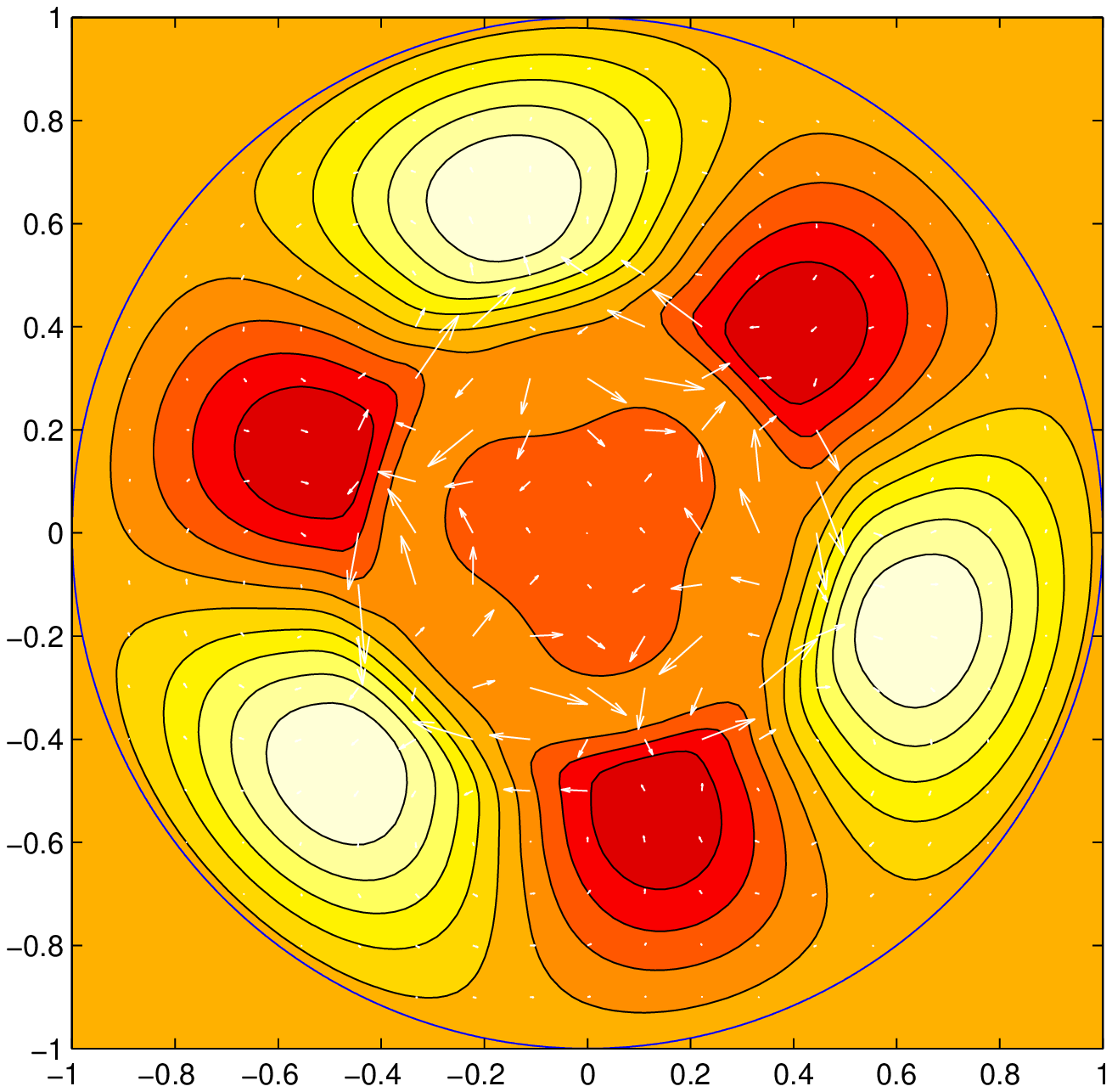, height=32mm}
   \end{tabular}
   \begin{tabular}{c}
      ({\it b}) \\
      \psfrag{Re}{$\Rey$}
      \psfrag{Ek/E0}{$ $}
      \epsfig{figure=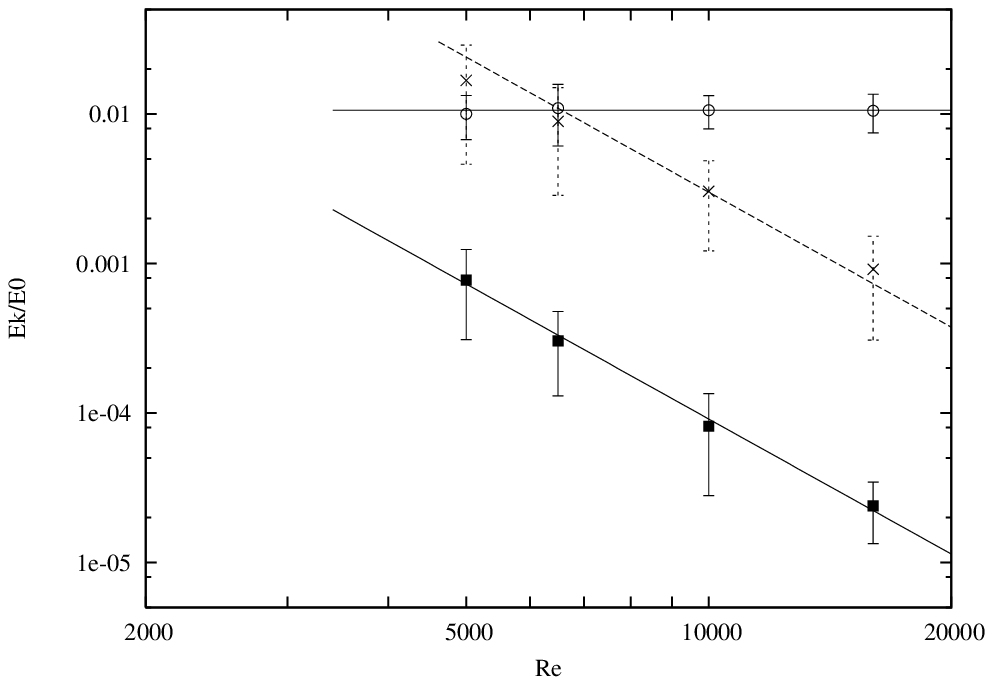, height=64mm}
   \end{tabular}
   \caption{\label{fig:slices}
      ({\it a})
      Cross-sections of an exact travelling-wave solution
      $L=\pi\,D$, $\Rey=6500$, wavespeed $c=1.563\,U$
      at $z=0,\frac{1}{2}L$.
      Dark regions are slower than the laminar
      profile and white regions faster.
      ({\it b})
      Normalised energy  $E'_{k\ne 0}/E_0$ on the boundary (squares)
      and of turbulence (circles). The crosses indicate the normalised
      total disturbance energy $E'/E_0$ on the boundary.
      The slope for the total disturbance energy (and
      also $E'_{k\ne 0}/E_0$ ) is $-3$,
      implying that the disturbance amplitude,
      $A\sim\Rey^{-1.5}$.  At $\Rey=5300$, $E'_{k\ne 0}/E_0\approx 0.011$
      in the reduced model and  $\approx 0.014$ in the fully 3D test case.
   }
\end{figure}

At the lowest $\Rey$ shown in figure \ref{fig:alp1edge} the energy
of the edge is highly variable in time. At the next $\Rey=6500$ a
period of slow variation is observed. \cite{duguet07} have recently
demonstrated that the edge can be used to find exact solutions by
identifying phases where the flow has a relatively simple temporal
behaviour. Such an episode is marked by a triangle where the
instantaneous flow was used as an initial condition for a
Newton--Krylov code. This converged to the exact solution pictured
in figure \ref{fig:slices}({\it a}), confirming that travelling wave
solutions do exist for the model.
The flow exhibits fast streaks towards the walls and slow streaks
are shed towards the centre. The state found is also very similar to
that marked by a circle at $\Rey=10\,000$ when the trajectory for a
while appears to settle towards a steady (translating) state. As
$\Rey$ increases the variability on the edge surprisingly decreases,
more so than can be explained by the increasing viscous time. The
viscous time $D^2/\nu$ scales as $\Rey$ in our time units, and
longer trajectories at larger $\Rey$ have been shown in figure
\ref{fig:alp1edge} to compensate. A possible explanation for the
decreasing variability is that the eigenvalues of the unstable
directions from the travelling waves decrease with increasing
$\Rey$
as in plane Couette flow \citep{viswanath07,wang07},
thus enabling closer and longer visitations.

The variability of the boundary in terms of energy or any other
chosen amplitude measure also, of course, indicates the range of
such attributes for  initial conditions which will trigger
turbulence. The fact that it is easier to trigger turbulence as
$\Rey$ increases
\citep[e.g.][]{hof03}
is reflected in the decrease in
the mean boundary energy as $\Rey$ increases seen in figure
\ref{fig:slices}({\it b}). Error bars in this figure represent the
mean and one standard deviation above and below this mean but
nevertheless, a clean scaling emerges for the disturbance amplitude of
$A\sim\Rey^{-1.5}$. \cite{mellibovsky06} found this scaling when
considering streamwise perturbations, and \cite{peixinho07} in
laboratory experiments with obliquely oriented jets.

\begin{figure}
   \begin{center}
      ({\it a})
      \epsfig{figure=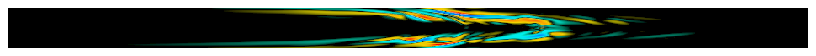, width=134mm} \\
      \epsfig{figure=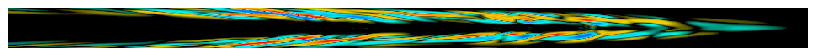, width=134mm} \\[5pt]
      ({\it b})
      \epsfig{figure=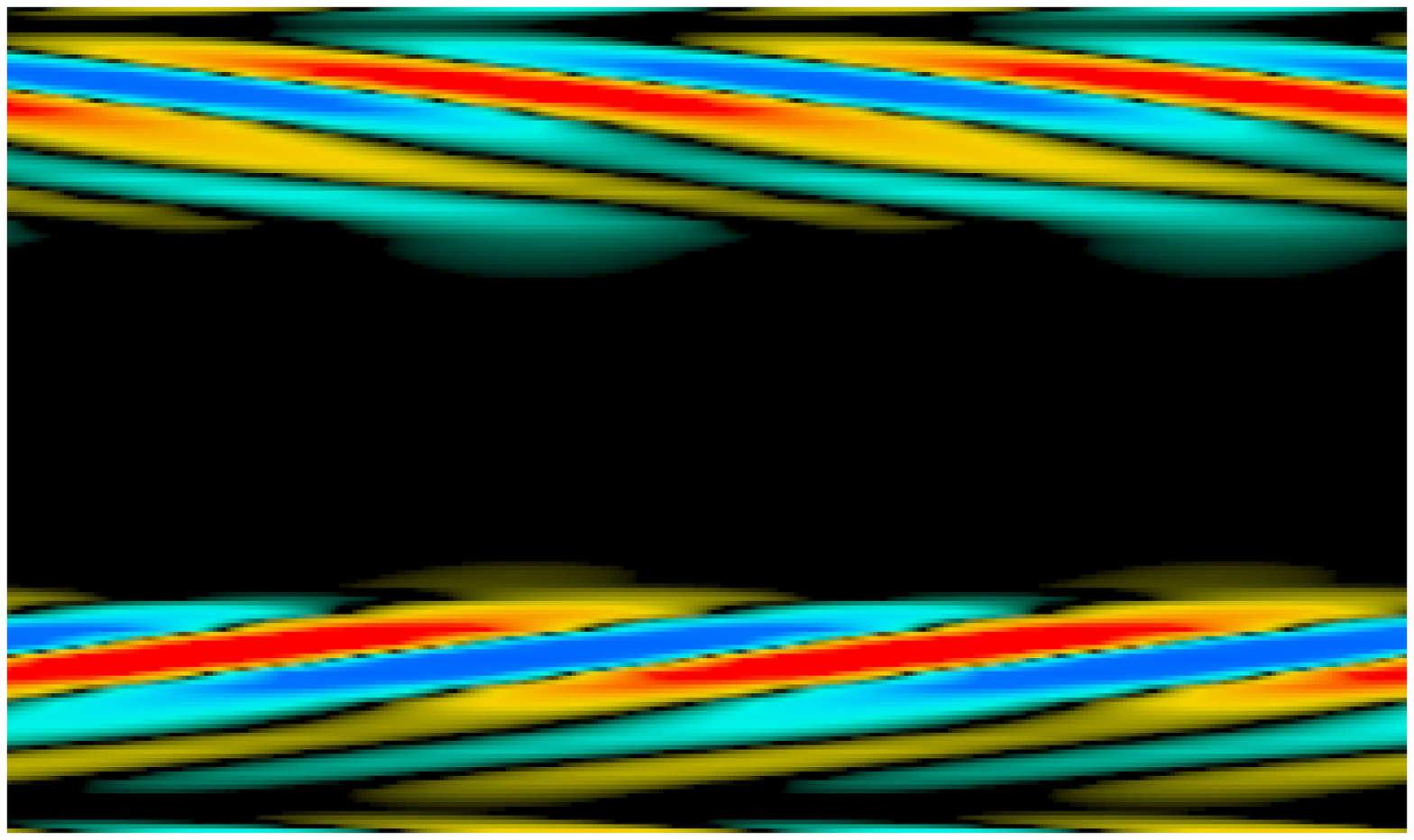, scale=0.12}
      \hspace{20mm}
      ({\it c}) \epsfig{figure=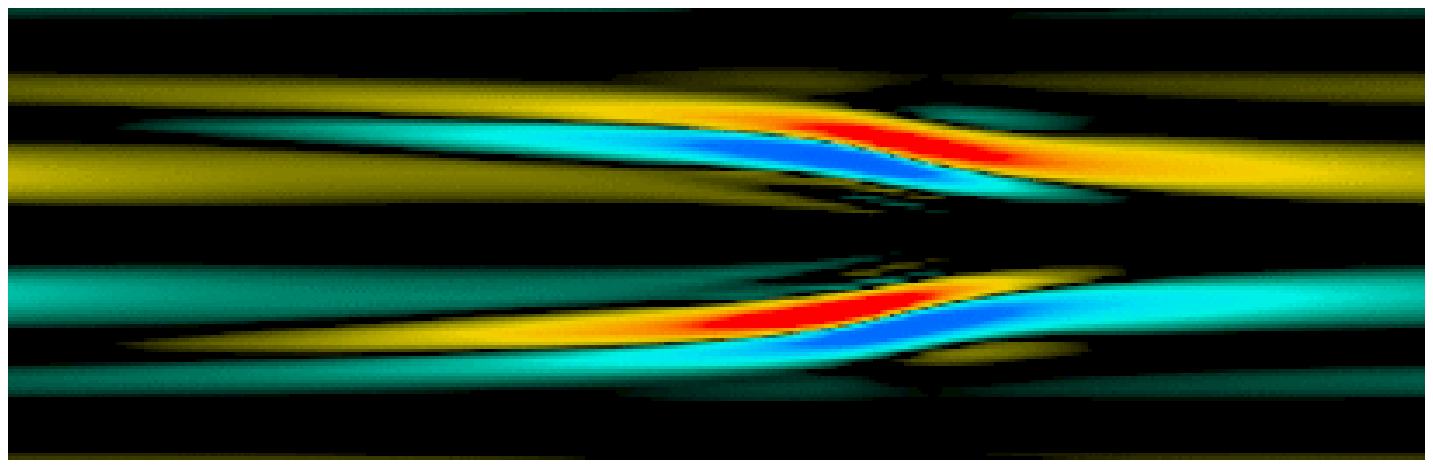, scale=0.25}
   \end{center}
   \caption{\label{fig:edgelong}
      ({\it a}) Snapshots of axial vorticity at two different times
      from a laminar-turbulent boundary trajectory.
      $L=16\pi\,D$, $\Rey=4000$. $20\,D$ of $\approx 50\,D$ shown.
      The disturbance is localised but with an extended upstream region.
      ({\it b})  An exact solution shown
      over two axial periods, $L=1.694\,D$, discovered using the
      code of \cite{pringle07}.
      ({\it c})  The exact solution of figure \ref{fig:slices}
      found by the Newton-Krylov code of \cite{duguet07}, $L=\pi\,D$.
   }
\end{figure}

\subsection{Characteristics of the laminar--turbulent boundary: long pipe}

The bisection procedure \citep{itano01,schneider07b,duguet07} for
isolating the laminar-turbulent boundary can equally be applied to
flow within a long pipe although the computational demands becomes
increasingly intensive. The 2$+\epsilon$-dimensional model is ideal for
a reconnaissance of likely behaviour, including identification of the
form the edge state takes.  With this motivation, trajectories on the
laminar-turbulent boundary were found for a pipe $\approx 50$
diameters long using the pressure gradient to distinguish between
edge and turbulent flow states. Interestingly, the attracting edge
state which emerges looks exactly like a turbulent puff except at
the trailing edge (upstream region): compare figures
 \ref{fig:puffslug} and  \ref{fig:edgelong}({\em a}). In a turbulent
puff this is the most energetic part possessing a fluctuation energy
level comparable to homogenised slug turbulence (see
\cite{willis08}) at higher $\Rey$. However, the edge state is
noticeably smoother even at $\Rey=4000$ (figure \ref{fig:edgelong})
compared to a puff at $\Rey=2600$ (figure \ref{fig:puffslug}) which
has finer scales. The similarities between the two are the strong
wall structures slanting into the axis at the upstream region, a
region where the axial vorticity reaches the axis (the trailing edge
region), followed by a gradual relaminarisation/decay downstream.
Interestingly it is the upstream region that waxes and wanes
(compare the two axial vorticity snapshots of figure
\ref{fig:edgelong}) rather than the passive-looking downstream wake.

The edge state remains localised just like the turbulent puff up to
$\Rey=3000$, but surprisingly also remains localised for much higher
$\Rey$ when the puff has given way to (global) slug turbulence. To
emphasise this, the localised edge state shown in figure
\ref{fig:edgelong} at $\Rey=4000$ was generated starting from the
global disturbance of figure \ref{fig:puffslug}. The fact that the
edge state remains puff-like throughout the puff-to-slug
transitional $\Rey$ range suggests that slug turbulence is
destabilised puff-turbulence rather than being a separate state
occupying a different part of phase space. This is certainly
consistent with simulations in which a puff state smoothly evolves
into a slug by slowly expanding upstream as well as downstream. A
corollary of this, of course, is that a puff and a slug cannot
coexist at a given $\Rey$: to our knowledge there are no reported
experimental observations to contradict this claim.

Figure \ref{fig:edgelong} also shows the axial vorticity for a
travelling wave with shift-and-reflect symmetry, found using the
method of \cite{pringle07} and that for the travelling wave shown in
figure \ref{fig:slices} (which has no special symmetry). Both have
axial vorticity slanted from the wall into the central axis
reminiscent of the turbulent puff and edge state. A recent search
for coherent fast-streak states within turbulent puffs has indicated
that the flow transiently resembles travelling wave states upstream
and downstream of the trailing edge region \citep{willis08}. Given
that the energetic trailing edge region is absent in the edge state,
there seems an even higher likelihood of seeing coherent states
there.

Although the findings in this 2$+\epsilon$-dimensional model are only
suggestive of what may occur in the fully 3-dimensional setting, they are
sufficiently interesting to motivate a fully 3-dimensional long-pipe
computation.  This is currently underway.

\section{Discussion}

In this paper, we have described the numerical formulation used to
simulate transitional pipe flow \citep{willis07,willis08,duguet07}.
This is based upon the poloidal-toroidal potential decomposition of
the velocity field discussed by \cite{marques90}, where the
difficulty of coupled boundary conditions has been by-passed by
influence matrix methods here. Reducing the system to five simple
second-order equations, the method is accurate, relatively simple to
implement and computationally efficient. This has then been used
here to explore dynamical subspaces in the hope of finding a reduced
system to aid understanding.

No evidence for turbulent transients has been found in axisymmetric
pipe flow, confirming an earlier investigation \citep{patera81}, or
in helical pipe flow consistent with the work by
\cite{landman90,landman90b}. Rotating pipe flow displays a classic
supercritical bifurcation route to turbulence
\citep{mackrodt76,toplosky88,landman90,landman90b,barnes00} which
has no bearing on the non-rotating situation. A brief search in
rotating helical pipe flow failed to find evidence for any
disconnected subcritical branches of solution which may have reached
back to the non-rotating limit. A 2$+\epsilon$-dimensional model, which
represents a minimal 3-dimensionalisation of the axisymmetric limit,
does, however, possess a subcritical transition scenario and all the
important spatio-temporal characteristics of fully resolved pipe
flow at $\Rey$ of the same order of magnitude. Localised
disturbances, structurally similar to turbulent puffs, are found in
the model at low $\Rey$ ($\approx 2600$), which slowly delocalise
at intermediate $\Rey$ ($\approx 3200$), and rapidly expand into
slugs at high $\Rey$ ($\approx 4000$). Exact unstable travelling
wave solutions also exist within the model and appear to underpin
the dynamics in phase space.

Within this 2$+\epsilon$-dimensional model, the relaminarisation
statistics of the puffs have been examined in pipes of varying
lengths. For pipes long enough to allow localised turbulence to
manifest itself, a critical $\Rey$ is suggested by the data above
which the `puff' becomes sustained.  On the other hand, if the pipe
is short so that turbulence fills the whole domain, the data is
consistent with only transient behaviour, i.e. the turbulence always
dies eventually. This qualitative change in behaviour as the
computational domain is varied is consistent with the seemingly
contradictory results found recently in fully 3-dimensional
simulations \citep{hof06,willis07}. The need to resolve the spatial
inhomogeneity of the puff state properly in numerical experiments is
clear.

Reconciling the conclusions drawn from the experimental data sets 
\citep{peixinho06,hof06} remains a challenge.  
The existence of both types of $\tau-\Rey$ scalings are not mutually 
exclusive, however, and resolution of the issue may be related to
the known sensitivity of the flow to the exact structure and 
amplitude of a perturbation. 
Experiments \citep{darbyshire95} and simulations 
\citep{faisst04,moehlis04} have indicated that the
laminar-turbulent boundary has a fractal-like structure
where very small changes to the perturbation can completely change
the fate of the flow.
Very small changes to the perturbation can completely change
the fate of the flow; finite lifetimes can then exist in the 
presence of an attractor
because an initial disturbance, apparently large enough to trigger a puff,
may actually not be within its basin of attraction 
\citep[e.g.\ ][]{moehlis04,mullin06}.
In addition, if the laminar-turbulent boundary is closely intertwined 
with the attractor in phase space, a trajectory can easily be nudged 
out of the attractor by noise effects, such as pipe roughness, 
temperature changes, pipe misalignment and vibrations.
Noise-induced relaminarisations of established puffs for 
$\Rey$ above $\Rey_c$ do not therefore contradict the existence of 
a critical $\Rey$.  
Conversely, noise could also artificially maintain puff turbulence 
in the absence of an attractor. There is clearly a need for further 
experimentation.

It is worth remarking that longer transients in the $100\,D$ pipe
(i.e. $\Rey$ closer to $\Rey_c$) could, in principle, have been
calculated given the computational savings available in the model.
It was found, however, that the puffs
begin to delocalise for $\Rey \approx 3200$ indicating that
by $\Rey_c=3450$, the puff has become unstable to slug-like
turbulence.  The same issue occurs in the real system:
puffs delocalise to become slugs for
 $\Rey=2250$-$2500$ \citep{willis08}. No claim has been made in
the literature that expanding slugs are other than permanently
sustained once generated.

The 2$+\epsilon$-dimensional model has also presented an opportunity to
probe the possible dynamics on the laminar-turbulent boundary
in long pipe flow. Calculations indicate that the attracting state
in this set is a localised puff-like structure which is smoother and
less energetic in the trailing edge region than its turbulent puff
counterpart. Also intriguingly, this end state remains localised way
beyond in $\Rey$ when the puff has delocalised. This tends to
suggest that the turbulent puff still exists as a solution  but has
become unstable to a slug state. The variability of the flow on the
laminar-turbulent boundary also highlights the variability in
initial conditions which can trigger turbulence. Just focusing on
the mean energy gives a disturbance amplitude scaling $A\sim\Rey^{-1.5}$
consistent with some numerical computations \citep{mellibovsky06} in
a short pipe and laboratory experiments with a carefully specified
jet configuration, designed to excite a coherent vortex \citep{peixinho07}.
Clearly, exploring how far this realisation can be
usefully developed is a promising area for future research.

In conclusion, we have introduced a model of pipe flow severely
truncated in its azimuthal degrees of freedom but otherwise fully
resolved in the others. This notwithstanding, the remaining system
captures all of the rich dynamical behaviour observed in pipe flow
but obtained at a fraction of the computational cost for the full 3-dimensional
situation. It therefore presents a very accessible arena in which to
test ideas and gain some insight quickly before deciding to invest a
considerable effort in the full 3-dimensional system.

\vspace{1cm}
\noindent
Acknowledgements: Many thanks to Yohann Duguet
and Chris Pringle for finding exact solutions in the model.
The authors would also like to thank an anonymous referee for
suggesting the $2+\epsilon$ nomenclature.
This research was funded by the EPSRC under grant GR/S76144/01.

%
%
%
%
%
%

\appendix
\section{Boundary conditions}
\label{app:BCs}
The Navier--Stokes equation (\ref{eq:govnonD}) plus the boundary condition
$\vec{u}=\vec{g}(\theta,z)$ are equivalent to (\ref{eq:govPot})
provided that on the boundary
\begin{equation}
   \label{eq:intBC}
   \vechat{n} \cdot \curl \left[
   (\pd{t} - \frac{1}{\Rey}\bnabla^2)\vec{u} + \vec{b}
   \right] = 0 ,
\end{equation}
where $\vechat{n}$ is its normal \citep[see][]{marques90}. This
condition ensures the term in the square brackets is equal to a
gradient, such as the pressure. If not imposed, this term may be any
$\chi\vechat{z}$ where $\laplace_h\chi=0$.  If this is not a gradient,
then an unknown body force is introduced.
For the axisymmetric case $\chi$ is
constant, the curl of $\chi\vechat{z}$ is then zero, and therefore
the condition is redundant.
Otherwise, from the diffusion term in (\ref{eq:intBC}), using the properties
$\bnabla^2\vec{u}=-\curl\curl\vec{u}$
and
$\curl\curl(\vechat{z}f) = \grad(\pd{z}f) - \vechat{z}\laplace f$,
one finds
\begin{eqnarray*}
   \vechat{r}\cdot\curl\bnabla^2\vec{u}
   & = & \vechat{r}\cdot\curl\curl(\vechat{z}\laplace_h\psi)
   + \vechat{r}\cdot\curl\curl\curl(\vechat{z}\laplace_h\phi)
   + \pd{zz}\vechat{r}\cdot\curl\vec{u} \\
   & = & \pd{rz}\psi_1 - \frac{1}{r}\pd{\theta}\phi_2
   + \pd{zz}\vechat{r}\cdot\curl\vec{g} ,
\end{eqnarray*}
which leads to the last condition of (\ref{eq:bcs}). The other
conditions simply express $\vec{u}=\vec{g}(\theta,z)$
with the gauge freedom $\phi=0$ on the boundary.

The simplified system for the axially averaged flow,
$h(r,\theta)$, arises because there exists a closed circuit $c$ that is not
simply-connected running along the
axial direction.  This has an associated condition
\[
   \int_c \left[
   (\pd{t} - \frac{1}{\Rey}\bnabla^2)\vec{u} + \vec{b}
   \right] \cdot \mathrm{d}l = 0, \nonumber
\]
which together with $P_z$ on the second curl of the Navier--Stokes
equations leads to the governing equation for $h$.  The boundary
condition $h=0$ assumes $P_z(g_z) = 0$, which otherwise would
correspond to a translating pipe or moving frame.

\section{Solution for coupled boundary conditions}
\label{app:reformulated}
Each Fourier mode for $\psi$,$\phi$ is expanded as the superposition
\begin{eqnarray}
   \psi(r) & = & \bar{\psi}(r) + a \, \psi^H(r) , \\
   {\phi}(r) & = & \bar{\phi}(r) + b \, \phi^H(r) ,
   \nonumber
\end{eqnarray}
where the coefficients $a$ and $b$ are scalars.
Subscripts $k$ and $m$ have been dropped.
The barred and superscripted functions solve two distinct systems.
Firstly,
\begin{eqnarray}
   (\pd{t} - \textstyle{\frac{1}{\Rey}}\laplace)\, \bar{\phi}_2
   & = &  - (1 - P_z) \, \vechat{z}\cdot\curl\curl\vec{b} , \\
   \laplace \bar{\phi}_1 & = & \bar{\phi}_2 , \nonumber \\
   \laplace_h \bar{\phi} & = & \bar{\phi}_1 , \nonumber \\
   (\pd{t} - \textstyle{\frac{1}{\Rey}}\laplace)\, \bar{\psi}_1
   & = & \vechat{z}\cdot\curl\vec{b} ,  \\
   \laplace_h \bar{\psi} & = & \bar{\psi}_1 , \nonumber
\end{eqnarray}
with boundary conditions
\begin{equation}
  \bar{\phi}_2 = \bar{\phi} = 0, \quad
  -\bar{\phi}_1 = g_z, \quad
  \pd{r}\bar{\psi}_1 = 0, \quad
  \left\{
     \begin{array}{ll}
        \bar{\psi} = 0 & \mbox{ if }m=0 \\
        -\pd{r}\bar{\psi} = g_\theta & \mbox{ if }m\neq 0
     \end{array}
  \right. ,
\end{equation}
where
$\laplace  \equiv (1/r)\pd{r}+\pd{rr}-m^2/r^2-\alpha^2k^2$ and
$\laplace_h\equiv (1/r)\pd{r}+\pd{rr}-m^2/r^2$.
This time-dependent system is written in matrix-vector form, according
to the time and radial discretisation, then inverted sequentially for
$\bar{\phi}_2 \to \bar{\phi}_1 \to \bar{\phi}$ and
$\bar{\psi}_1 \to \bar{\psi}$.
The second homogenised system is
\begin{eqnarray}
   (\pd{t} - \textstyle{\frac{1}{\Rey}}\laplace)\, \phi_2^H
   & = &  0 , \\
   \laplace \phi_1^H & = & \phi_2^H , \nonumber \\
   \laplace_h \phi^H & = & \phi_1^H , \nonumber \\
   (\pd{t} - \textstyle{\frac{1}{\Rey}}\laplace)\, \psi_1^H
   & = & 0 , \\
   \laplace_h \psi^H & = & \psi_1^H , \nonumber
\end{eqnarray}
with boundary conditions
\begin{equation}
  \phi_2^H = 1, \quad \phi_1^H = \phi^H = 0, \quad
  \pd{r}\psi_1^H = 1, \quad
  \left\{
     \begin{array}{ll}
        \psi^H = 0 & \mbox{ if }m=0 \\
        \pd{r}\psi^H = 0 & \mbox{ if }m\neq 0
     \end{array}
  \right. .
\end{equation}
As this system has no time-dependent $\vec{b}$, solutions with
superscript $H$ may be precomputed. The original boundary conditions
on $\phi$ and $\psi$ are satisfied upon reconstruction from the
barred and superscripted variables. Two boundary conditions are
satisfied by construction for all cases: as $\bar{\phi}$ and
$\phi^H$ satisfy trivial boundary conditions, then ${\phi}=
\bar{\phi} + b \, \phi^H =0$ on the boundary;  similarly
$-\phi_1=g_z$ is satisfied
automatically.

For axisymmetric modes, $m=0$, the system is of lower order as $\psi$
always appears as $\pd{r}\psi$, including in the boundary condition.
The condition involving $\vec{b}$ is not required for this case
(see appendix \ref{app:BCs}).
The simplest solution is to add
the boundary condition $\psi=0$ so that we may invert for
$\psi$ for all modes.
When $m=0$ the remaining two boundary conditions
which couple the potentials are satisfied by selecting
scalars $a$ and $b$ according to the
following evaluated on the boundary:
\begin{equation}
   \quad a = -(\pd{r}\bar{\psi}+g_\theta)/\pd{r}\psi^H,
   \quad b = -(\pd{rz}\bar{\phi}-g_r)/\pd{rz}\phi^H
   \quad
   \mbox{ if }m = 0 .
\end{equation}

For non-axisymmetric modes, $m\neq 0$, the condition $-\pd{r}\psi=g_\theta$ is
satisfied automatically.  The last two conditions are satisfied by
solving the system for $a$ and $b$ evaluated
on the boundary,
\begin{equation}
   \left[
      \begin{array}{ll}
         \frac{1}{r}\pd{\theta}\psi^H &  \pd{rz}\phi^H \\
         -\pd{rz}\psi_1^H & \frac{1}{r}\pd{\theta}\phi_2^H
      \end{array}
   \right]
   \left[
      \begin{array}{ll}
        a \\ b
      \end{array}
   \right]
   \, = \, - \,
   \left[
      \begin{array}{ll}
        \frac{1}{r}\,\pd{\theta}\bar{\psi} + \pd{rz}\bar{\phi} - g_r\\
        \Rey\, \vechat{r}\cdot\curl\vec{b} - \pd{zz}\,\vechat{r}\cdot\curl\vec{g}
      \end{array}
   \right]
   \quad \mbox{ if }m\neq 0.
\end{equation}
As this only requires the inversion of a 2$\times$2 matrix and the
$H$-functions are pre-computed, this is an inexpensive way to ensure
all boundary conditions are simultaneously satisfied to machine
precision.

\bibliographystyle{jfm}
\bibliography{trans}


\end{document}